\documentclass[12pt]{iopart} 
\pdfoutput=1
\usepackage{graphicx} 
\usepackage[caption=false]{subfig}
\usepackage{color}
\usepackage{float,amssymb}

\usepackage{lineno}
\usepackage{setspace}

\begin{document} 

\title[]{ Exploring the interplay between small and large scales 
movements 
in a neotropical small mammal
}

\author{E. Brigatti$^{\star}$, B. R\'ios-Uzeda$^{\ddag}$, M. V. Vieira$^{\ddag}$}
\address{$^{\star}$ Instituto de F\'{\i}sica, Universidade Federal do Rio de Janeiro, 
Av. Athos da Silveira Ramos, 149,
Cidade Universit\'aria, 21941-972, Rio de Janeiro, RJ, Brasil}
\address{$^{\ddag}$ Laborat\'orio de Vertebrados, Instituto de Biologia, Universidade Federal do Rio de Janeiro,
Caixa Postal 68020, 21941-590, Rio de Janeiro, RJ, Brasil }
\ead{edgardo@if.ufrj.br}


\onehalfspacing



\newpage

\begin{abstract} 

We record and analyze the movement patterns of the marsupial  {\it Didelphis aurita}
at different temporal scales.
Animals trajectories are collected at a daily scale by using spool-and-line techniques, 
and  with the help of  radio-tracking devices 
animals traveled distances are estimated at intervals of weeks.
Small-scale movements are well described by truncated L\'evy flight,
while large-scale movements produce a distribution of distances 
which is compatible with a Brownian motion.
A model of the 
movement behavior of these animals, based on a truncated L\'evy flight calibrated
on the small scale data, converges towards a Brownian 
behavior after a short time interval of the order of one week. 
 
These results show that
whether 
L\'evy flight or Brownian motion behaviors
apply, will depend on the scale of aggregation of the animals paths. 
In this specific case, as the effect of the rude truncation present in the daily data generates
a fast convergence towards Brownian behaviors, L\'evy flights become of scarce interest 
for describing the local dispersion properties of these animals, 
which result 
well approximated by a normal diffusion process and not a fast, anomalous one. 
Interestingly, we are able to describe two movement phases 
as the consequence 
of a statistical effect generated by aggregation,
without the necessity 
of introducing ecological constraints
or mechanisms operating at different spatio-temporal scales.
This result is of general interest, as it can be a key element for describing  movement phenomenology at distinct spatio-temporal scales across different taxa and in a variety of systems.

\end{abstract}






\maketitle

\section{Introduction}





In the last two decades the analysis of a large amount of biological data 
related to animal displacements has shown that the 
distribution of the 
movement single steps
can be well characterized by the use of heavy tailed distributions \cite{16,32,92,59,93,95,95b}.
Even if some concerns about these results have been raised \cite{45}, and 
the risk of describing such a variety of behaviors and
strategies with an oversimplified theoretical framework can exist,
nowadays it seems evident that stochastic processes 
which can generate such type of behaviors
must be included in the indispensable 
toolbox for modeling animal movements.
These processes, generally known as L\'evy processes \cite{levy1},
present some specific traits which 
 are 
 very different from those of ordinary Brownian processes.
In fact, they produce
trajectories with a peculiar geometry,
marked by abrupt long jumps that connect 
clusters of frequent short displacement,
which generates a characteristic scale-invariant structure.
Moreover, such L\'evy processes trigger 
super-diffusive behavior \cite{26,49,metzler},
which means that the mean squared displacements of individuals are a super-linear function of time
and the limiting distributions of the total traveled distances present power law behavior.
These features lead to an efficient way of exploring the space, in contrast to the slowness 
of standard diffusive behavior.
In fact, L\'evy flights have been suggested to be the best option
for solving the problem of random search \cite{16,16b},
and, for this reason, 
the optimal choice for describing
searching and foraging movements,
specifically for
environments with homogeneous and scarce resources \cite{30}. 

L\'evy flights, similarly to ordinary random walks, can be defined as a 
sum of independent identically distributed random steps.
In this case, the step distributions present power law tails
which generate divergent variances. 
For this reason, the central limit theorem does not apply,
and the limiting distributions  of the total traveled distances 
present power law behaviors \cite{metzler}.\\

The fact that the single step distributions
present infinite variance can discourage their use, as 
in ecological systems an unavoidable cutoff is always present.
In fact, the step movement of an individual is evidently limited
because of natural, physical and energetic constraints.
Moreover, a realistic animal movement is characterized by physiological and circadian rhythms which cause direct limits on the maximum step length.
For this reason, the power law behaviour can hold only in a limited spatial range,
and some type of cutoff must be considered.
These facts suggest the use of truncated power laws \cite{89,45}, an important 
element taken into account in different studies \cite{45,30,mashanova,rhee,boris,rosario}.

 
The presence of the cutoff means that the variance is finite and, in principle, it should cause the 
convergence to a Gaussian behaviour. 
But this convergence, in general, is slow and it is reached only after a very large number of steps \cite{rosario}.
In this conditions, the distribution of the total traveled distances maintain statistical properties indistinguishable from the ones generated by L\'evy flights. Only increasing the number of realized steps, a crossover between L\'evy and Gaussian regimes is observed, which depends on  the cutoff and the exponent values of the L\'evy flights \cite{rosario}. Therefore, an open question is when such large number of steps are achieved in the movements of specific organisms, causing a change from a L\'evy to a Gaussian regime.

These theoretical results have been registered in some empirical studies.
Two distinct modes in the walking behaviour of aphids was recorded by Mashanova {\it et al.} \cite{mashanova}.
Similarly, in the case of human mobility, the heavy-tailed 
behaviours have been reported within a certain region, but not on a global scale \cite{rhee,gonzalez}.
The relationship between L\'evy and Gaussian regimes in movement paths could therefore depend on the 
scale of the tracking data. The same organism could present a L\'evy regime at fine-scale movements that could change to Gaussian as the interval between locations increases.
This interesting question remains still open, as empirical studies which describe the movements of the same species at pronounced different scales are sparse \cite{Fryxell,Torney,Kazimierski}.
Fryxell {\it et al.} \cite{Fryxell} studied elks over different orders of magnitude in time (minutes to years) and space (meters to 100 km).  This study is based on descriptive statistics and qualitatively shows and characterizes elk shifts from dispersive to home-ranging phases. 
 At the finest scales, elk used area-restricted search while browsing, and less sinuous paths when not browsing.
The study by  Kazimierski {\it et al.} \cite{Kazimierski} studied the movements of a small american marsupial qualitatively characterizing the shape of the small scales trajectories and estimating the speed of movement, the daily home range (less than $100$ $m^2$) and its exploration by using the large-scale dataset.
.\\


The main purpose of our study is to characterize the movement patterns of a 
Neotropical marsupial at different temporal scales.
In a first experimental set, animal trajectories were mapped continuously and in detail, 
during a tracking that lasts up to 
one night of activity (ca. 8h).
These data were used to estimate the single step distribution $P(\ell)$, where the step 
$\ell$ is defined as the distance between two successive turning points.
In a second field work, based on radio-tracking locations,
positions of animals were recorded at larger time intervals, 
of the order of weeks.
Animals were released in the same site of capture,
and their location was determined every night, for several nights.
From these measurements it was possible to estimate
the total traveled distances distribution $P(r)$, where $r$ is the 
geographical distance traversed in a week time.
The statistical properties of these two distributions were analysed, showing
the presence of a truncated L\'evy flight for $P(\ell)$, with parameters that generate 
a total traveled distances distribution $P(r)$ with a Brownian character, for large 
temporal scales. The number of steps necessary for reaching such
distribution is estimated, allowing a consistency check between
the distinct movement characteristics present at different scales.

\section{Materials}

These experiments were conducted in the Iconha and Guapi-Macacu river basin, which are part of the municipalities of Guapimirim and Cachoeiras de Macacu, in the Rio de Janeiro State, Brazil.

For describing the features of animals movements at different temporal scales we use two complementary approaches: with spool-and-line devices and with radiotracking. 
The experiments of the first set, with spool-and-line devices, were conducted in the Serra dos \'Org\~aos National Park, PARNASO, the largest continuous Atlantic Forest remnant in the State of Rio de Janeiro (20,020 ha), southeastern Brazil ($22.479^{\circ}$ S, $42.988^{\circ}$ W, datum WGS84), between 500-700m altitude. The forest is part of the montane rainforest complex, in an old-growth successional stage (details in \cite{Finotti20}).

The experiments of the second set, with radiotracking, were conducted in a nearby location, in forest fragments varying from $5$ to less than $10000$ ha area, immersed in a matrix of pastures near the Reserva Ecol\'ogica Guapia\c cu - REGUA ($22^{\circ}25'$ S, $42^{\circ}44'$W). 
Linear distances between opposing edges of forest fragments 
varied between ca. 200m to dozens of kilometers.
The original forest of these fragments was disturbed by different degrees of anthropogenic activities. Canopy is 20 m high approximately, but varying in continuity between forest fragments.
Subcanopy is fairly open (details of the structure of forest fragments in \cite{Finotti12}).

The first experiment  record movements 
of {\it Didelphis aurita}, a small marsupial released in the forest of PARNASO. 
Individuals were equipped with a spool-and-line device \cite{72}, which consisted 
in a bobinless cocoon, containing 
up to 480 m of thread, wrapped in a PVC film. 
This device was glue to the back of animals with synthetic resin, and the end of the thread was attached to a mechanism \cite{74} which marked the starting point of the animal path.  
Trajectories were tracked following the thread and measuring 
the azimuth angles  corresponding to a pronounced change in the direction of motion
(a deviation greater than $5\,^{\circ}$), and the distances between these turning points.
Note that the use of a spool-and-line device is a better replacement for GPS
tracking when, as in our case, the considered species lives in region of dense forest 
and intricate topography, which cause low reliability in GPS positioning. 

In the second experiment, performed in REGUA, individuals of {\it D. aurita} were equipped with a radio-tracking device, released at the same site of capture, and tracked for one night every week.
Individuals were captured-recaptured from 2014 to 2015 as part of a population ecology study in forest fragments near REGUA \cite{Uzeda2017}. A total of 14 {\it D. aurita} (subadults or adults of both sexes) were selected for tracking based on good physical condition.
The chosen individuals were taken to the laboratory of the REGUA
located in the study area, and kept in a plastic cage (41 x 34 x 18 cm) for no more than 24 h, with ad libitum food and water, and covered to reduce disturbance and stress. 
Early in this 24h period the individual had a radiotag tracking device with an active sensor (VHF radio tracking, Telenax inc.) attached on its back at a point just below the scapulas. The radiotag was fixed with flexible and water-resistant glue (Super Bonder Power Flex, LoctiteTM), and ranged between 4 and 10 g, depending on the animal weight, represented a maximum five percent of its total weight. After receiving the radiotag, the individual remained under observation overnight to detect any apparent disturbances in behavior, and to evaluated if the radiotag was fixed and working properly. 
Animals were released the next day on the same site of capture. 
After release, they 
were tracked for a whole night once a week using the homing-in technique \cite{White1990}. As marsupials are nocturnal, monitoring took place mainly during the night, when signal location was determined every two hours. A receiver TR-4 and an antenna RA-14K (Telonics, Mesa, USA) were used to detect and close in the signal of individuals until we were sure the animal was less than 5 m away, based on the quality and intensity of the signal. Only when an individual moved more than 20 m a new location was georeferenced. Individuals were monitored until their radiotag fell, the animal died, or the signal was lost for two consecutive weeks. 
From these data, the geographical distances traversed in a week time were extracted.

\begin{figure}[h]
\centering
\includegraphics[width=0.8\textwidth, angle=0]{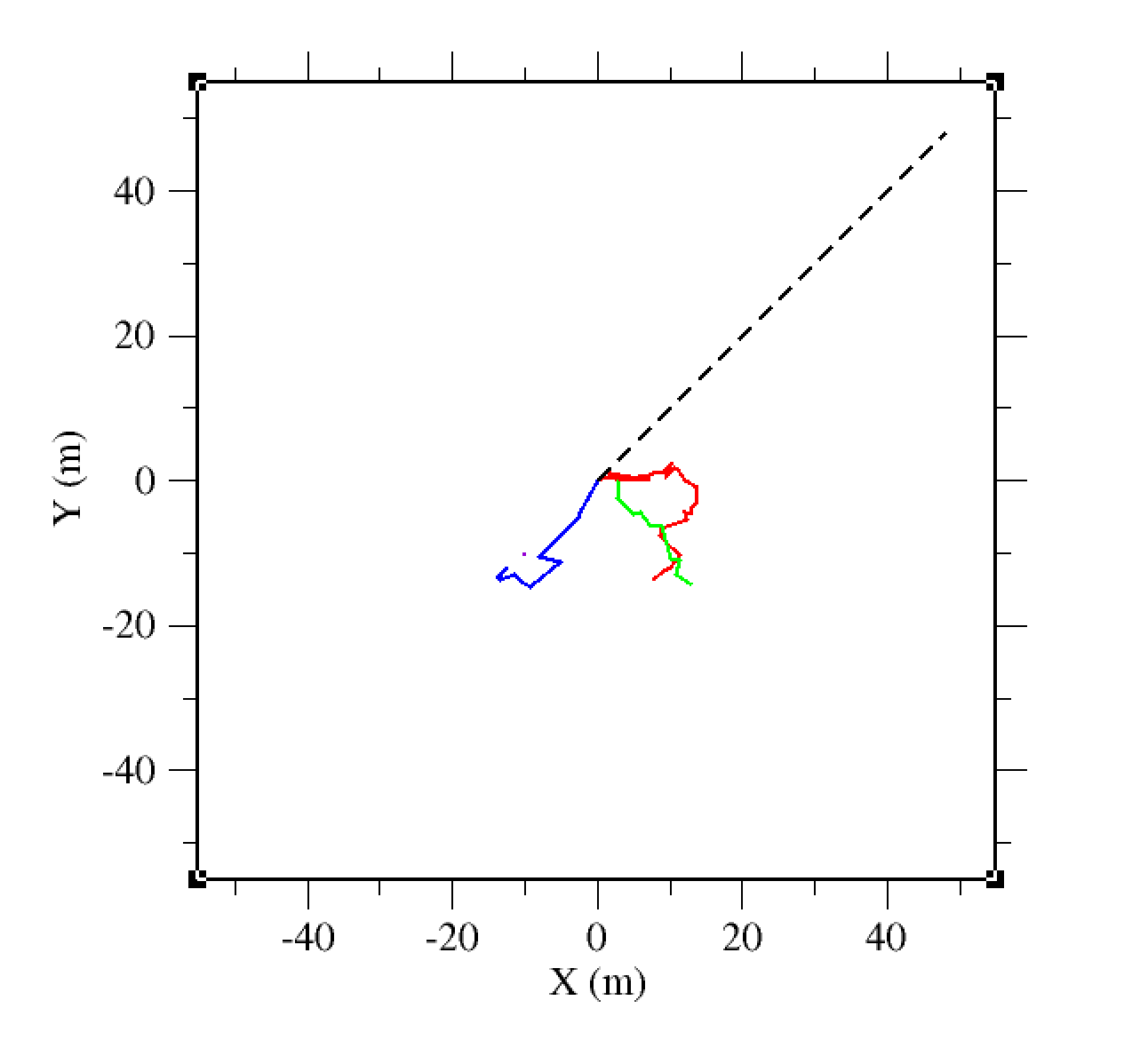}
\caption{Colored continuous curves represent some trajectories 
tracked at small scales with the spool-and-line technique. 
The black dashed straight line represents the mean distance travelled
by an individual after one week, 
as estimated from the dataset obtained using radiotelemetry techniques. 
The typical animals size is comparable with the width
of this line. 
 }
\label{fig:paths}
\end{figure}

A graphic representation showing the typical spatial scales
covered by the two experiments is shown in Figure \ref{fig:paths}.

 
\section{Methods} 

\subsection{Small scales}

For the data collected during one night of activity, 
we quantify the distribution of the step-lengths $\ell$,
defined as the distances between two successive turning points
of a given trajectory.
The turning points can be identified as the points with a change in the direction of motion 
larger than a fixed minimal angle. 
However, such an ad-hoc discretization 
is arbitrary and the estimated distribution $P(\ell)$ would strongly depend on this choice \cite{23}.
We overcome this problem following a recent approach introduced in \cite{32}.
In analogy with that procedure, a trajectory 
is projected on one 
axis 
and the projected step 
is defined  as the distance between two inversions in the direction of the movement of the
projected trajectory. 
It was analytically proven that in the presence of a power law distribution
for the original path in 2D, the distribution for the projected paths preserves the 
same power law relationship \cite{32}. 
In the case of an exponential distribution, the projected data 
maintain the exponentially decaying behaviour.
Following this method, we are able to rigorously investigate the shape of the 
$P(\ell)$ distribution using a segmentation procedure which
does not depend on any arbitrary choice.
As pointed out in \cite{32,23}, there is only an important 
detail to be taken into account. 
The operation of projection causes the proliferation of spurious data 
corresponding to steps similar or even smaller than the smallest
step-length presents in the original dataset. 
These data must be excluded from the final analysis.
Moreover, as the last measured step-length can have a length influenced 
by the ending of the spool, 
they were eliminated from the data set too.\\

Once obtained $P(\ell)$, we look for 
the best supported 
model for describing our data. 
We are interested in distinguish if 
the distributions present power law or
exponential behaviors.
For the power law behaviour we consider the 
Pareto distribution:
$f(x)=(\mu-1)a^{\mu-1}x^{-\mu}$, for $x\ge a$,
and the truncated Pareto distribution, defined as:
$g(x)=(1-\mu)/(b^{1-\mu}-a^{1-\mu})x^{-\mu}$, for $a \le x\le b$.
Finally, the exponential distribution is defined as:
$h(x)=\lambda\exp(\lambda a) \exp(-\lambda x)$, for $x\ge a$.
As already commented in the introduction,
the use of a truncated Pareto distribution is essential.
In fact, 
the direct physical limits characteristic of a realistic animal movement,
produce inevitable truncation effects.
For this reason, it is important to consider an upper limit 
for the $\ell$ values.


The best estimation of the parameters of these distributions can be 
obtained using the Maximum Likelihood Estimation (MLE) technique.
This technique has shown to be the most 
reliable compared to classical least squares methods \cite{17,89,90}.
For the truncated Pareto distribution, the estimation of the $\mu$ parameter  must be realized 
numerically; details can be found in \cite{90,91,boris}.


Finally, a quantitative selection between the three considered fitting models
is obtained using
the Akaike information criterion (AIC), which compares models likelihoods, penalizing models with more parameters \cite{Anderson}.
The AIC is defined as: $AIC=2K-2L$.
$L$ corresponds to the maximum log-likelihood,
which is estimated following Edwards \cite{45}; $K$ is the number of parameters of the model, which is 2 for the exponential and Pareto distribution and 3 for the truncated Pareto.
The best supported
model is the one which displays the lowest AIC.

\subsection{Large scales}

The data collected during the radiotracking field study
were filtered, considering only movements of individuals
which 
do not leave the fragment where they where first detected. 
In fact we considered only the movement behavior of settled animals, disregarding animals that left the forest fragment for dispersal.
We measured the geographical distance between the first animal position (the vector $\bar x_1$)
and the secondary location ($\bar x_2$), measured one week later: $r=|\bar x_2-\bar x_1|$.
From these data it is possible to quantify the frequency of traveled distances of individuals 
as a function of geographical distance $r$.
Our aim is to distinguish if the data are better described by a 
classical distribution expected for a normal diffusive behaviour,
or by a distribution generated by L\'evy walks.  
The first situation corresponds to 
a Rayleigh distribution, which is the one expected 
for 
a classical Brownian process. 
The distribution has the following aspect:

\begin{equation}
P(r)=\frac{r}{\sigma^2} \exp(-r^2/2\sigma^2),
\label{eq:Rayleigh}
\end{equation}

and the best estimation of its single parameter is 
obtained using the Maximum Likelihood Estimation technique.
A simple calculation gives: 
$\sigma_{MLE}=[\frac{1}{n}\sum^n_{i=1}r_i^{2}/2]^{1/2}$,
where $n$ is the sample size, and its approximate variance is: $\sigma_{MLE}^2/4n$.


For testing if the data are well described by distributions generated by L\'evy walks,
we should consider the family of L-stable distributions, characterized by power-law tails \cite{metzler}.
Unfortunately, most members of the L-stable family 
have no closed form and they are characterized by four parameters.
For these reasons, they are not well suited for a 
direct goodness-of-fit test.
However, in order to consider the L-stable family as an alternative model, 
the crucial aspect is to demonstrate 
that data present a clear power law upper tail.
In fact, it is the tail 
the most important portion of the distribution for determining if
the central limit theorem applies or not.
For this reason, it is reasonable to focus on determining 
how heavy is the upper tail of the distribution. 
If we assume that our distribution decays as $F(x)\propto x^{-\alpha}$ for $x>M$, 
where $M$ is the median of the distribution and $\alpha>0$,
the distribution is heavy-tailed if $\alpha<3$.
The estimation of the tail index $\alpha$ is obtained by using
the conditioned maximum likelihood estimator 
introduced by Hill \cite{hill}. 

This approach has some important advantages over 
popular models comparison  or specific test based on general distributions.
In fact, distributions with two power law tails are 
used in the literature \cite{general}, but, despite  
they may present some convenience in the statistical description of the dataset, these general probability distributions do not have any plausible theoretical
connection with movement models.  
Moreover, in general, a comprehensive fitting of the data has the greatest descriptive power 
for the central region, 
but it says less about the tails of the distribution. 


\subsection{Exploring the connection between small and large temporal scales}

Based on the results of the analysis 
of the data at small and large scales, we
will explore the connections between
these two regimes. Our aim is to test the robustness of our results 
and to clarify the relationships linking the two scales.
In particular, we want to show how it is possible
to obtain the principal statistical 
features of the large scale distribution as an emergent behavior generated by
the properties of the small scale walks.
For achieving these goals, we implement some
simulations, based on discrete walks, for synthetical reproducing the 
empirical 
$P(r)$ distribution. 
The discrete walks are parametrized using the results obtained from the analysis of the 
small scale data of {\it Didelphis aurita}. 
Fixed the number of steps $n$, 
for each walk, we calculate the related traveled distance $r_i$. 
By generating a large amount of walks  we can
obtain the distribution $P_{n}(r)$.
In dependence of $n$, this distribution presents very different characters.
In fact, a small $n$ generates
a distribution with a well defined heavy upper tail; increasing the $n$ value 
produces a shape change towards a typical Rayleigh distribution. 
This evolution in the shape of the distribution
can be described by measuring  the tail index $\alpha$ defined previously. 
Alternatively, we can measure the maximum of the log-likelihood of the  dataset of the $r_i$, 
which are generated by the simulations, 
considering a Rayleigh distribution ($L(r_i |\sigma_{MLE})$) 
and the maximum of the same log-likelihood of 
a dataset of equal size, but composed by independent, identically-distributed 
draws ($d_i$) from a Rayleigh distribution with the same parameter $\sigma$ ($L(d_i |\sigma_{MLE})$). 
For our setting, the ratio $R=\frac{L(r_i |\sigma_{MLE})}{L(d_i |\sigma_{MLE})}$, 
will be  very close to one  
when the distribution of the $r_i$ is  practically represented by a Rayleigh distribution, and it will decrease 
towards zero the more the distribution $P_{n}(r)$ departs from this shape.
Note that we do not use the logarithm of the likelihood-ratio
but the ratio of the log-likelihoods because 
its numerical output can more easily provide a rough description of the similarity between the two distributions.

Finally, we will shed light on the diffusion properties of these 
stochastic walks
depicting the dependency of the typical value of these distributions on $n$.

\section{Results}

\subsection{Small scales}

\begin{figure}[h]
\centering
\includegraphics[width=0.8\textwidth, angle=0]{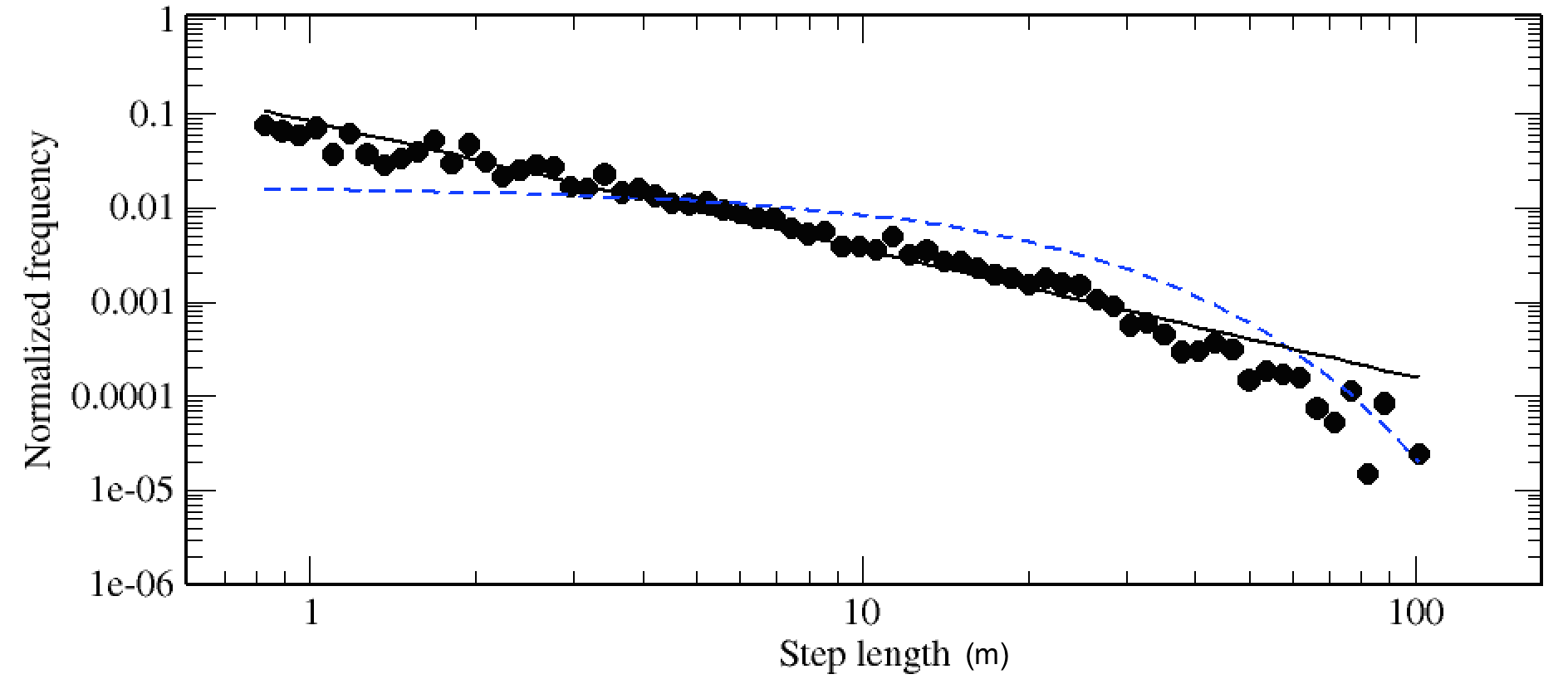}
\caption{Log-binned data with the best
estimated Pareto-truncated distribution (continuous lines)  and exponential distribution (dashed line).
Step-lengths are expressed in meters.
Analyzed data consider steps coming from the projections onto the two axis ($x$ and $y$) together.
In fact, as expected, the two projections present indistinguishable behaviors.
 }
\label{fig:fitting}
\end{figure}

We analysed 141 different trajectories,
where each trajectory is the whole track of a single individual movement. 
These data generated a sample size of 2239 steps, which
corresponds to the number of regressed data after eliminating the spurious $l$.
The calculation of the AIC values shows that the Pareto-truncated distribution
presents the lowest AIC, which indicates that
it is the best model for the description of the data (see Table \ref{table0}).
This fact can be visually 
appreciated in Figure \ref{fig:fitting}, where we plot the  
data with the best estimated model. 
The Pareto-truncated distribution obtained 
using the MLE approach 
is quite satisfactory. 
The exponent $\mu$ of the Pareto-truncated distribution is equal to
$1.36\pm0.02$ and the parameter $a$ and $b$ are 
respectively $0.8$ and $103.1$ meters.

\begin{table}
    \centering
{\small

\begin{tabular}{|c|c|c|c|c|}  
\hline
 & Pareto-truncated & Pareto & Exponential  \\
\hline
\hline
Loglikelihood  &  -6712 & -6937  & -6965 \\
\hline
AIC  & 13431 & 13879 &  13934  \\
\hline
$\Delta AIC$  &  0  & 448  &  503 \\
\hline
$wAIC$  & 1 & 0  & 0 \\
\hline 
\hline
\end{tabular}
}
\caption{ Models selection between Pareto-truncated, Pareto and Exponential distribution. 
$\Delta AIC$ is the difference of the AIC values and $wAIC$ is the weighted value of the $AIC$ 
of the considered models.
}
\label{table0}
\end{table}
 
\subsection{Large scales}

\begin{figure}[h]
\centering
\includegraphics[width=0.8\textwidth, angle=0]{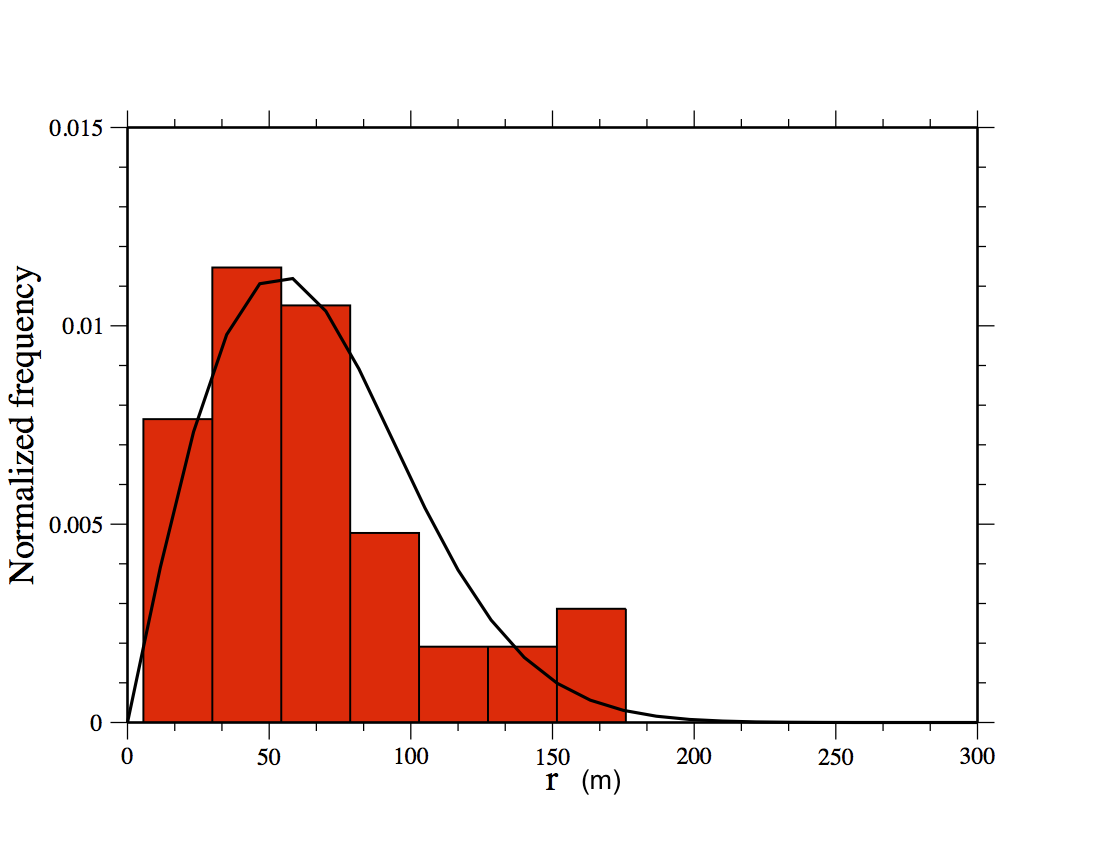}
\caption{Histogram of $P(r)$ as obtained from the telemetry measurements, with the best
estimated Rayleigh distribution (continuous line). The sample size is 43, which corresponds to the number of one-week distances measured, combining data from all individuals.
}
\label{fig:ray}
\end{figure}

The histogram of the collected data with the best estimated Rayleigh distribution
is plotted in Figure \ref{fig:ray}.
A visual inspection suggests that data are well described by this model,
in particular considering that the Rayleigh distribution depends only on a 
single parameter.
The best estimated parameter obtained
by means of the MLE is:
$\sigma_{MLE}=  54\pm4$ $m$. 
A Kolmogorov-Smirnov test of our observations against 
the fitted Rayleigh distribution gives
a p-value of $0.166$, which suggests that we cannot reject the hypothesis that 
the field data come from the fitted distribution.

The estimation of the tail index $\alpha$, as obtained using the 
Hill's estimator, obviously depends on the fraction of observations 
used in the tail estimation. 
For every subset of observations which satisfies $r>M$, 
the estimated $\alpha$ is always greater than 3, 
and it increases with decreasing subsets, rejecting the hypothesis of a power law tail ($\alpha$ should be $< 3$).
In particular, 
a subset corresponding to the largest $58\%$
of the original dataset gives $\alpha=3.3\pm0.5$.
Diminishing the subset size the tail index increases,
reaching $4.1 \pm 0.9$ when the $30\%$
of the observations are used.


\subsection{Exploring the connection between small and large temporal scales}

\begin{figure}[h]
\centering
\includegraphics[width=0.75\textwidth, angle=0]{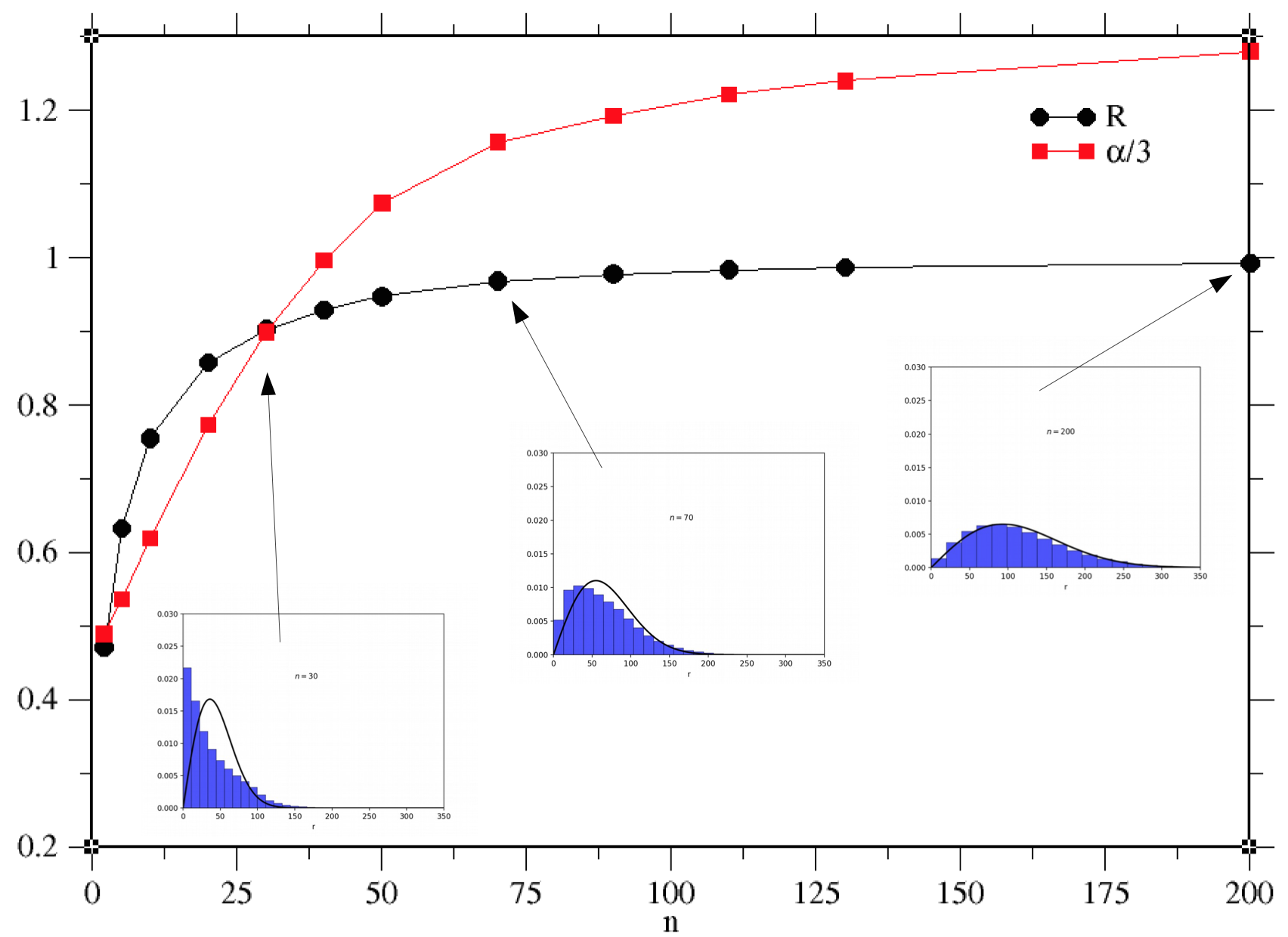}
\includegraphics[width=0.8\textwidth, angle=0]{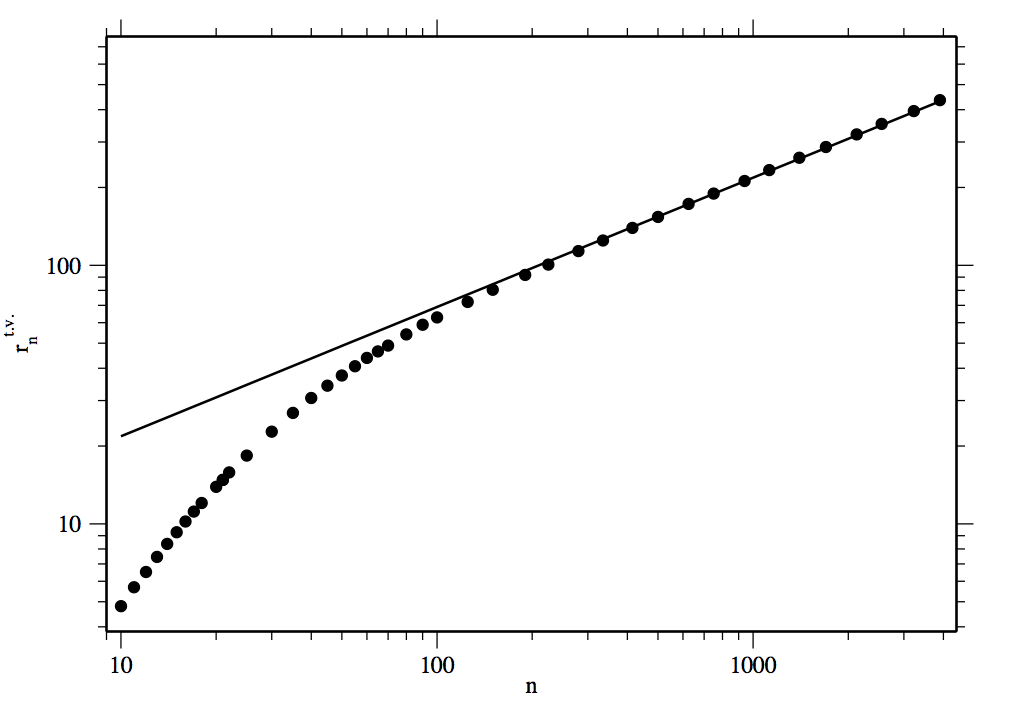}
\caption{Top: Tail index $\alpha$ normalized by 3, and $R$, as measured
from the different distributions $P_n(r)$. Continuous lines are just
a guide for the eyes.
In the inboxes, the histograms represent the $P_n(r)$ for $n=30,70,200$,
as obtained from the simulation of 200000 different walks.
The continuous lines are the corresponding best estimated Rayleigh distributions.
Bottom:  typical values of the $P_n(r)$ as a function of $n$. 
As a measure of typical value we use the geometric mean, which better characterizes 
the distributions when they are close to power laws. For large $n$ we can observe that $r^{t.v.}_n\sim n^{1/2}$ (continuous line).
}
\label{fig:X}
\end{figure}

Individual movements are simulated 
performing a discrete walk which implements
a Pareto truncated step-length distribution with isotropic
directions of motion.
Once fixed the number of steps $n$,
we measure the distance $r_i$
between the starting point and the final individual position. 
We run 200000 different simulations 
of the walk and we obtain the histogram of the $r_i$ for that $n$.
We can observe that by changing the value of $n$, 
these histograms 
undergo a change in their shape. 
When the number of performed 
steps is small, a power law tail is clearly recognizable, but increasing
the number of steps it disappears, generating a distribution
characterized by a Rayleigh-like shape. 
This phenomenon is well described in Figure \ref{fig:X},
where we plot the tail index $\alpha$ and $R$, as measured from 
the different distributions $P_n(r)$.
In this case, the $\alpha$ are measured selecting 
an upper-order statistics corresponding to the
$40\%$ of the original dataset.
This value is chosen because it produces
a consistent estimations of $\alpha$ for small $n$. 
For $n<40$, $\alpha$ is smaller than 3, which implies
the presence of heavy upper tails.
This fact finds a correspondence in the 
small $R$ values, which suggests
that a Rayleigh distribution is a poor description of the data.
In contrast, for $n>40$, $\alpha$ is larger than 3, revealing
the disappearance of the power law tail.  The $R$ value comes close to
1, suggesting that now the Rayleigh distribution can well 
approximate our data. For values of $n$ larger than 100,
effectively $R\approx 1$ and $P_n(r)$ becomes indistinguishable from
a Rayleigh distribution.

These simulation data can also characterize, in a clear and intuitive way, 
the diffusion properties of these walks.
In Figure \ref{fig:X}  we display the typical value $r^{t.v.}_n$ of the $P_n(r)$ as a function of $n$.
For $n$ large enough, 
the behaviour predicted by the central limit theorem,  
where $r^{t.v.}_n\sim n^{1/2}$, is recovered. 
In this regime $r$ is the sum of a sufficient large number of $l$ and the
maximum values reached by these $l$ 
are directly determined by the cut-off value $b$. 
For small $n$, we can observe 
a superdiffusive behaviour ($r^{t.v.}_n\sim n^{k}$, with $k>1/2$).
In this regime the number of sorted step lengths $l$ is reduced. It follows that the maximum values 
reached by these $l$ are generally small compared to the cut-off and its effect can be partially neglected.
If we could effectively forget the cut-off and consider the distributions as  power laws with exponent 
$\delta$, the typical values should grow as $n^{1/(\delta-1)}$ \cite{metzler}.
As in our case the cut-off is very sharp and the $\mu$ exponent is relatively close to $1$, even if a power law can be detected, it is difficult to robustly estimate the corresponding exponent.
The transition between the two regimes is quite smooth and
the progressive convergence towards the normal diffusion set in for $n$ close to 100, in accordance with the previous analysis of the $P_n(r)$ shape.

\begin{figure}[h]
\centering
\includegraphics[width=0.8\textwidth, angle=0]{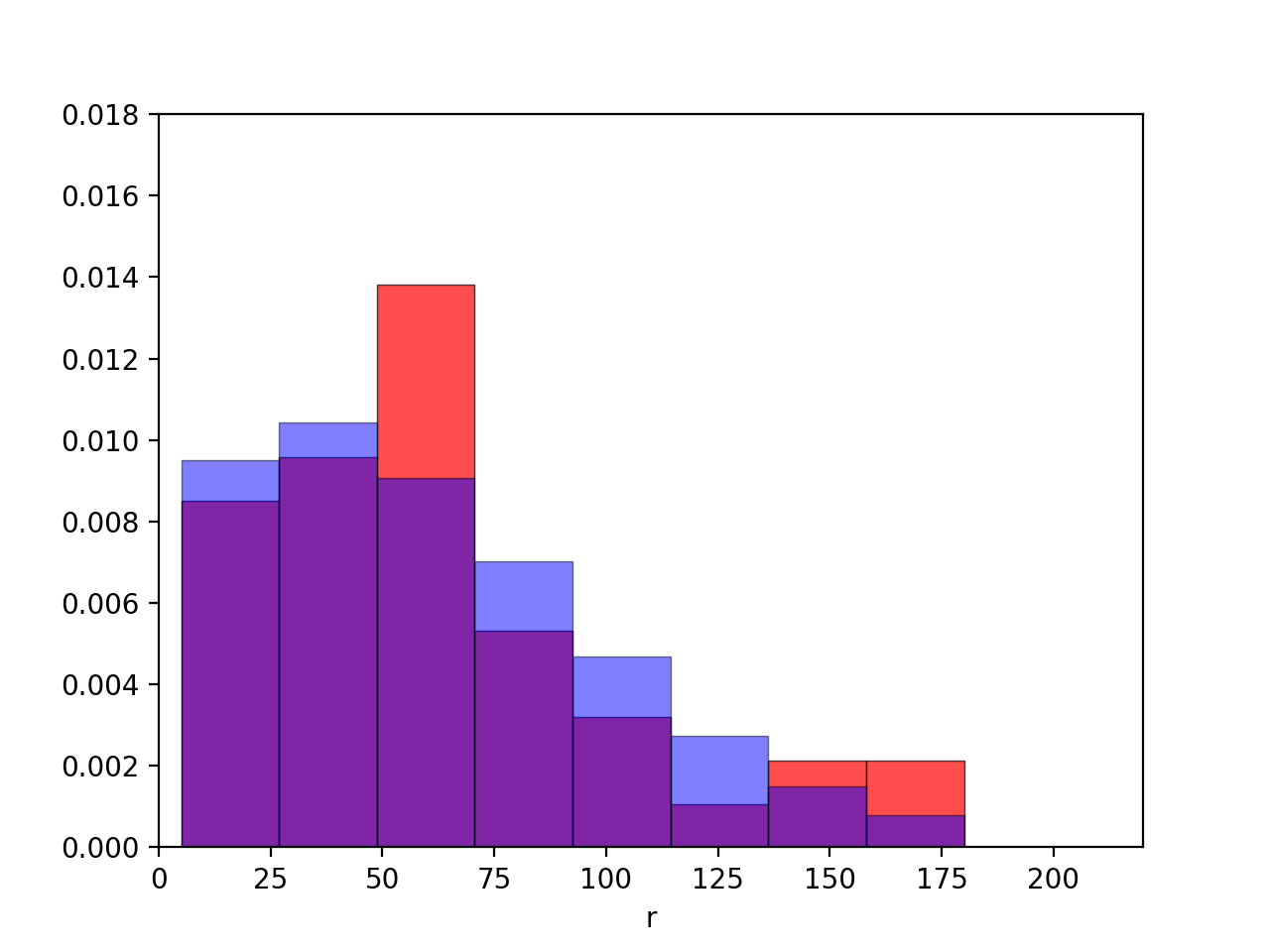}
\caption{The blue histogram represents the distances $r$ generated by a simulation
which uses a Pareto truncated distribution with the parameters calibrated from
 the  empirical dataset when it implements walks of 68 steps. 
Data are obtained from 200000 different walks.
The red histogram illustrates the field data as obtained from the telemetry measurements.
}
\label{fig:simul}
\end{figure}

Finally, we estimated the best $n$ value
which reproduces the $P(r)$ of the field work dataset.
This is determined by looking for the $\sigma_{MLE}$
value which best approximates the one of the empirical data.
For a number of steps fixed to 68, $\sigma_{MLE}=54$, exactly the same of the empirical case. 
In Figure \ref{fig:simul}, it is possible to appreciate 
how the histograms of the simulated and experimental dataset
are effectively very similar.






\section{Discussion}



In the literature it is common to find works which describe the geometry of 
paths followed by different species of animals, characterized by a variety of body sizes and, conversely, geographical scales.
In general, these analysis collect data on short temporal scales (hours or daily scales)
and use different specific methods of tracing and segmentation of the paths.
When large temporal scales are considered, 
studies are more interested in describing dispersion properties,
and, for illustrating long-distance migration and dispersal, other methodologies, 
like biotelemetry techniques and satellite tagging, are typically used.
In contrast, analyses which focus on a single species and describe movements at all relevant scales,
relating the small to the large ones, are few \cite{Fryxell,Torney,Kazimierski}.
These studies are very important 
for the development of the movement ecology paradigm \cite{Nathan08}, 
which presents the goal of unifying organismal movement research also from the perspective of temporal
scales: from a single step toward the entire lifetime track, eventually characterizing the presence of different
movement phases.
In this work we have explored these ideas from a particular point of view.
We do not emphasize the importance of different biological processes, but we just
scrutinize the statistical effects generated on movements by 
crossing through different temporal scales,
without introducing any 
scale-dependent ecological effect or mechanism.

In order to fully describe the statistics of general movements, in principle, it would be 
necessary to typify the distribution of the movement steps at every time scale;
in fact, the distribution of the daily steps could be different from the weekly or monthly ones.
In simplified cases, it is possible to reconstruct the distributions corresponding to 
large time scales only knowing the one describing the short time scales.
In fact, from a theoretical perspective, 
if the short time scale distributions are Gaussians or L\'evy ones,
the situation is particularly favorable.
In this case, as these distributions are stable,
at all time scales the step distributions will be described by the same stable law,
with their parameters values appropriately tuned.
This well known result can be generalized to the case when
movement steps are summed from independent distributions with finite variance. 
Then, the central limit theorem assures that
the distribution describing long times steps converges towards a Gaussian, if 
the number of steps is sufficiently hight.
Depending on the rate of convergence, this result can be of greater or lesser interest,
and, for example, 
a L\'evy flight with truncation can lead to Gaussian behaviors.

These theoretical results can be applied to real-world data, 
implying that whether a truncated L\'evy flight or a Brownian motion 
apply will depend on the scale of aggregation
of the paths of the considered animals.
By resampling at smaller scales and retesting the
model at each scale, discernible truncated L\'evy or Brownian 
modes may come into focus at the appropriate scale. 
In our case, at the scale of one-day paths, a truncated L\'evy flight 
fits very well steps length data. 
Zooming out, however, we would expect a truncated L\'evy flight model to be 
falsified when the scale matches a sufficient large sum of steps. 
Indeed, our analysis shows that the L\'evy truncated distribution which characterizes the daily
movements of {\it 
Didelphis aurita} converges towards a Brownian model after
only one week. This means that the effect of the rude truncation present in the daily
data generates a fast rate of convergence towards Brownian behaviors.  
For this reason, L\'evy  flights become of scarce interest in 
describing the dispersion properties of these animals inside the considered fragments, 
which result described by a normal diffusion process, and not a fast, anomalous one.
This result has a clear ecological relevance, as normal diffusion 
processes produce shorter typical translation distances
than the anomalous ones, generated by L\'evy flights.
This implies a lower propensity to explore regions further away from the origin of the movement.

Our study presents some connections with the current growing interest in describing the crossover from superdiffusive to normal-diffusive dynamics in different particles systems such as, for example, in actively moving biological cells. However, with the exception of a study on aphids \cite{mashanova}, this phenomenon has not yet been quantitatively recorded in the case of animal movements.
Note that this crossover is generally characterized by looking at the  temporal dependence of the mean squared displacement. A linear dependence corresponds to the normal  diffusion law; super-linear relations, in the form of a power-law, correspond to superdiffusive behaviors. In our study we are not able to describe this phenomenon by looking directly at the mean squared displacement since our trajectories did not present temporal tracking and did not entirely span the considered time scales.

Finally, it is important to note that the effect of summation of enough different steps is sufficient 
for generating the Rayleigh distribution. 
Here it is not necessary to introduce other biological or ecological constraints, such as
fragment size, aversion to densely occupied regions, 
feeding spots, and home range,  
nor to typify a variety of ecological processes operating at diverse spatio-temporal scales, 
for producing different movement phases \cite{Nathan08}. 
In this way we are able to predict  movements over large temporal scales
through simulations calibrated from 
empirical results obtained over short time periods \cite{Revilla04,Erika}, 
without the necessity of introducing other effects. 
Of course,
ecological constraints on movement behavior do exist, in general and for {\it Didelphis aurita} in particular \cite{72,74,Uzeda2017}. Home range area and location is frequently considered one 
such ecological constraint, that could limit superdiffusive movement, generating subdiffusive or normal-like diffusion (review in \cite{Benhamou2014}). However, home range actually is 
more a consequence of interactions between individual behavior and their environment, 
not an environmental factor itself. 
Our results suggest that factors determining home 
ranges are more likely to be affecting short term responses, manifest in the distributions of step lengths and turning angles. Long term normal-like diffusion could be just a consequence of these short-term effects. 
\\

This study shows how very simple models like traditional discrete walks may be adequate for describing 
movement data at many temporal resolutions, producing results that are not obvious, but can be easily interpreted.
Even when a L\'evy flight results not to be appropriate for describing 
large scales behaviors, it provides a reference point for a parsimonious explanation of the
overall process of movement.
In this sense a L\'evy flight can be seen, together with the Brownian motion, as an appropriate 
null model, which acts as a 
more useful starting point than other more elaborated models.

L\'evy processes have been accepted as  appropriate models describing movement paths of a variety of organisms \cite{Reynolds 2018}, but conflicting views remains on the processes they represent. 
Some argue that they are mostly limited to describe movements at large spatial scales, and L\'evy processes would be a result of a series of random walk processes acting at different spatial scales \cite{Benhamou2014,Pyke2015}. 
Conversely, truncated L\'evy processes may  approximate random walks, or Brownian motion, as the number of steps increases between points where positions are registered \cite{rosario}. 
Our results presents evidence that the latter occurs in nature, in the paths of the marsupial {\it Didelphis aurita}. 
More than one mechanism could result in a L\'evy process \cite{Reynolds 2018}. The realization of this connection between L\'evy and Brownian regimes in the modeling of animal paths is essential to understand what mechanism is determinant in a particular system. 


\section*{Declarations}

\subsection*{Ethical Approval}

Trapping and handling conformed to guidelines sanctioned by the American Society of Mammalogists \cite{Sikes2016}, and was approved by IBAMA/MMA (Authorization numbers 87/05-RJ, 099/06-RJ, 13861-1, 13861-2, 16703).

\subsection*{Funding}
M.V.V. was supported by CNPq (grants 308.974/2015-8, 441.589/2016-2), and Funda\c c\~ao de Amparo \'a Pesquisa do Estado do Rio de Janeiro - FAPERJ (grant E-203.045/2017).
 
\subsection*{Availability of data and materials} 

If the paper is accepted for publication data will be permanently archived in a no-cost general repository option such as Figshare.

\ack

The author Boris R\'\i os Uzeda greatly contributed to the realization of this study, but sadly he was not able to see
the publication of this work, as he 
suddenly passed away. 
We always miss him as a kind colleague and dearest friend.



\newpage

\newpage

\newpage

\newpage

\end{document}